\RequirePackage{fix-cm}
\documentclass[smallextended]{svjour3}       % onecolumn (second format)
\smartqed  % flush right qed marks, e.g. at end of proof

\usepackage{graphicx}
\usepackage[top=30truemm,bottom=30truemm,left=25truemm,right=25truemm]{geometry}
\usepackage{color}
\usepackage{amsmath}
\usepackage{lscape}
\usepackage{amsfonts}
\usepackage{bm}
\usepackage{color}
\usepackage{here}
\usepackage[top=30truemm,bottom=30truemm,left=25truemm,right=25truemm]{geometry}
\graphicspath{{figures/}}

\definecolor{green}{rgb}{0.2, 0.55, 0.02}

\begin{document}

\title{Model order reduction with neural networks: Application to laminar and turbulent flows}


\author{Kai Fukami \and Kazuto Hasegawa \and Taichi Nakamura \and Masaki Morimoto \and Koji Fukagata %etc.
}


\institute{Kai Fukami \at
              Department of Mechanical and Aerospace Engineering, University of California, Los Angeles, CA 90095, USA\\
              Kazuto Hasegawa, Taichi Nakamura, Masaki Morimoto, Koji Fukagata \at
              Department of Mechanical Engineering, Keio University, Yokohama, 223-8522, Japan \\
              \email{kfukami1@g.ucla.edu}       %  \\
}

\date{Received: date / Accepted: date}

\maketitle

\begin{abstract}
{
We investigate the capability of neural network-based model order reduction, i.e., autoencoder (AE), for fluid flows.
As an example model, an AE which comprises of convolutional neural networks and multi-layer perceptrons is considered in this study.
The AE model is assessed with four canonical fluid flows, namely: (1) two-dimensional cylinder wake, (2) its transient process, (3) NOAA sea surface temperature, and (4) a cross-sectional field of turbulent channel flow, in terms of a number of latent modes, the choice of nonlinear activation functions, and the number of weights contained in the AE model.
We find that the AE models are sensitive to the choice of the aforementioned parameters depending on the target flows.
Finally, we foresee the extensional applications and perspectives of machine learning based order reduction for numerical and experimental studies in the fluid dynamics community.
}
\end{abstract}

\section{Introduction}

Thus far, modal analysis has played a significant role in understanding and investigating complex fluid flow phenomena.
In particular, the combination with linear-theory based data-driven and operator-driven tools, e.g., proper orthogonal decomposition (POD)~\cite{Lumely1967}, dynamic mode decomposition~\cite{Schmid2010}, and Resolvent analysis~\cite{MS2010}, has enabled us to examine the fluid flows with interpretable manner~\cite{TBDRCMSGTU2017,THBSDBDY2020}.
We are now able to see their great abilities in both numerical and experimental studies~\cite{noack2011reduced,NPM2005,bagheri2013koopman,LANOT2018,Schmid2011,YT2019,NFL2017}.
In addition to these efforts aided by linear-theory based methods, the use of neural networks (NNs) has recently been attracting increasing attentions as a promising tool to extract nonlinear dynamics of fluid flow phenomena~\cite{BNK2020,BEF2019,Kutz2017,duriez2017machine}.
In the present paper, we discuss the capability of the NN-based nonlinear model order reduction tool, i.e., autoencoder, while considering canonical fluid flow examples with several assessments.

In recent years, machine learning methods have exemplified their great potential in fluid dynamics, e.g., turbulence modeling~\cite{D2021,DIX2019,ling2016reynolds,gamahara2017searching,maulik2017neural,maulik2019sub,MSRV2019,YZWX2019,wu2018physics}, and spatio-temporal data estimation~\cite{FNKF2019,SP2019,CZXG2019,FFT2019a,FFT2019b,HLC2019,MFF2020,FFT2021b,morimoto2020generalization,erichson2019,MTMFF2021,MFZNF2021}.
Reduced order modeling (ROM) is no exception, referred to as machine-learning-based ROM (ML-ROM).
An extreme learning machine based ROM was presented by San and Maulik~\cite{SM2018}.
They applied the method for quasi-stationary geophysical turbulence and showed its robustness over a POD-based ROM.
Maulik et al.~\cite{MFRFT2020} applied a probabilistic neural network to obtain a temporal evolution of POD coefficients from the parameters of initial condition considering the shallow water equation and the NOAA sea surface temperature.
It was exhibited that the proposed method is able to predict the temporal dynamics while showing a confidence interval of its estimation.
For more complex turbulence, Srinivasan et al.~\cite{SGASV2019} demonstrated the capability of a long short-term memory (LSTM) for a nine-equation shear turbulent flow model.
In this way, the ML-ROM efforts combined with the traditional ROM are ongoing now.

Furthermore, the use of autoencoder (AE) has also notable role as an effective model-order reduction tool of fluid dynamics.
To the best of our knowledge, the first AE application to fluid dynamics is performed by Milano and Koumoutsakos~\cite{Milano2002}.
They compared the capability of the POD and a multi-layer perceptron 
{(MLP)-}based AE using the Burgers equation and a turbulent channel flow.
{
They confirmed that the linear MLP is equivalent to POD~\cite{BH1989} and demonstrated that the nonlinear MLP provides improved reconstruction and prediction capabilities for the velocity field at a small additional computational cost.
}
More recently, convolutional neural network (CNN)-based AEs have become popular thanks to the concept of filter sharing in CNN, which is suitable to handle high-dimensional fluid data set~\cite{FFT2020}.
Omata and Shirayama~\cite{omata2019} used the CNN-AE and POD to reduce a dimension of a wake behind a NACA0012 airfoil.
They investigated the difference of the temporal behavior in the low-dimensional latent space depending on the tool for order reduction
{
and demonstrated that the extracted latent variables properly represent the changes in the spatial structure of the unsteady flow field over time.
They also exhibited that the capability of this method to compare flow fields constructed using different conditions, e.g., different Reynolds numbers and angles of attack.
}
Hasegawa et al.~\cite{HFMF2019,HFMF2020a} proposed a CNN-LSTM based ROM to predict the temporal evolution of unsteady flows around a bluff body by following only the time series of low-dimensional latent space.
{
They consider flows around a bluff body whose shape is defined using a combination of trigonometric functions with random amplitudes and demonstrated that the trained ROM were able to reasonably predict flows around a body of unseen shapes, which implies generality of the low-dimensional features extracted in the latent space.
}
The method has also been extended to the examination of
Reynolds number dependence~\cite{HFMF2020b} and turbulent flows~\cite{nakamura2020extension},
{
whose results also unveiled some major challenges for CNN-AE, e.g., its capability against significantly different flow topologies from those used for training and the difficulty in handling very high-dimensional data.
}
From the perspective on understanding the latent modes obtained by {CNN-}AE, Murata et al.~\cite{MFF2019} suggested a customized {CNN-}AE referred to as a mode-decomposing CNN-AE which can visualize the contribution of each machine learning based mode.
{
They clearly demonstrated that, similar to the linear MLP mentioned above, 
the linear CNN-AE is equivalent to POD,
the nonlinear CNN-AE improves the reconstruction thanks to the nonlinear operations,
and a single CNN-AE mode represents multiple POD modes.
}
Apart from these, we can also see a wide range of AE studies for fluid flow analyses in~\cite{EMM2019,CJKMPW2019,xu2020multi,maulik2020reduced,LPBK2020}.
As a result of such eagerness for the uses of AE, we here arrive at the following open questions:
\begin{enumerate}
    \item How is the dependence on a lot of considerable parameters for AE-based order reduction, e.g., a number of latent modes, activation function, and laminar or turbulent?
    \item Where is the limitation of AE-based model order reduction for fluid flows?
    \item Can we expect further improvements of AE-based model order reduction with any well-designed methods?
\end{enumerate}
{
These questions should commonly appear in feature extraction using AE not only for fluid mechanics problems but also for any high-dimensional nonlinear dynamical systems.
To the best of authors' knowledge, however, 
the choice of parameters often relies on a trial-and-error,
and
a systematic study for answering these questions seems to be still lacking.
}

Our aim in the present paper is to demonstrate the example of assessments for AE-based order reduction in fluid dynamics, so as to tackle the aforementioned questions.
Especially, we consider the influence on (1) a number of latent modes, (2) choice of activation functions, and (3) the number of weights in an AE with four data sets which cover a wide range of nature from laminar to turbulent flows.
{Findings through the present paper regarding neural-network-based model order reduction for fluid flows can be applied to a wide range of problems in science and engineering, since fluid flows can be regarded as a representative example of complex nonlinear dynamical systems.
Moreover, because the demand for model order reduction techniques can be found in various scientific fields including robotics~\cite{saku2021spatio}, aerospace engineering~\cite{amsallem2008interpolation,brunton2021data}, and astronomy~\cite{cheng2020identifying}, we can expect that the present paper can promote the unification of computer science and nonlinear dynamical problems from the perspective of application side.}
Our presentation is organized as follows.
We introduce the present AE models with fundamental theories in section~\ref{sec:method}.
The information for the covered fluid flow data sets is provided in section~\ref{sec:data}.
We then present in section~\ref{sec:result} the assessments in terms of various parameters in the AE application to fluid flows.
At last, concluding remarks with outlook of AE and fluid flows are stated in section~\ref{sec:conclusion}.

\section{Methods and theories of autoencoder}
\label{sec:method}

\subsection{Machine learning schemes for construction of autoencoder}
\label{sec:AEscheme}

%%%%%%%%%%%%%%%%%%%%%%%%%%%%%%%%%%%%%%%%%%%%
\begin{figure}
	\vspace{0mm}
	\centering
		%\hspace{-30mm}
		\includegraphics[width=1.00\textwidth]{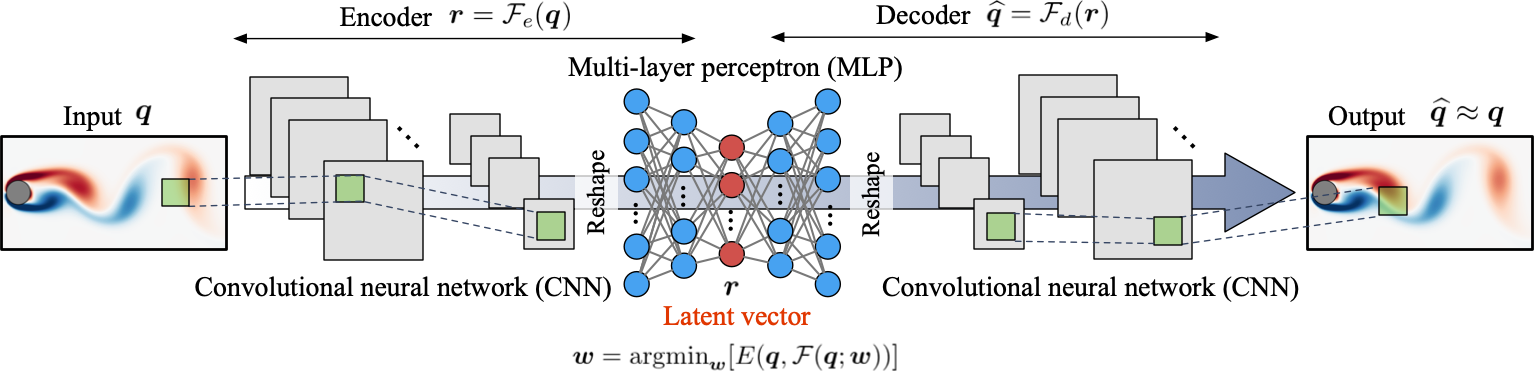}
		\caption{Autoencoder (AE) used in the present study. The present model is constructed by a combination of multi-layer perceptron and convolutional neural network.}
		\label{fig_ae}
\end{figure}
%%%%%%%%%%%%%%%%%%%%%%%%%%%%%%%%%%%%%%%%%%%%

In the present study, we use an autoencoder (AE) based on a combination of a convolutional neural network (CNN) and a multi-layer perceptron (MLP), as illustrated in figure~\ref{fig_ae}.
Our strategy for order reduction of fluid flows is that the CNN is first used to reduce the dimension into a reasonably small dimensional space, and the MLP is then employed to further map it into the latent (bottleneck) space, as explained in more detail later.
In this section, let us introduce the internal procedures in each machine learning model with mathematical expressions and conceptual figures.

%%%%%%%%%%%%%%%%%%%%%%%%%%%%%%%%%%%%%%%%%%%%
\begin{figure}
	\vspace{0mm}
	\centering
		%\hspace{-30mm}
		\includegraphics[width=1.00\textwidth]{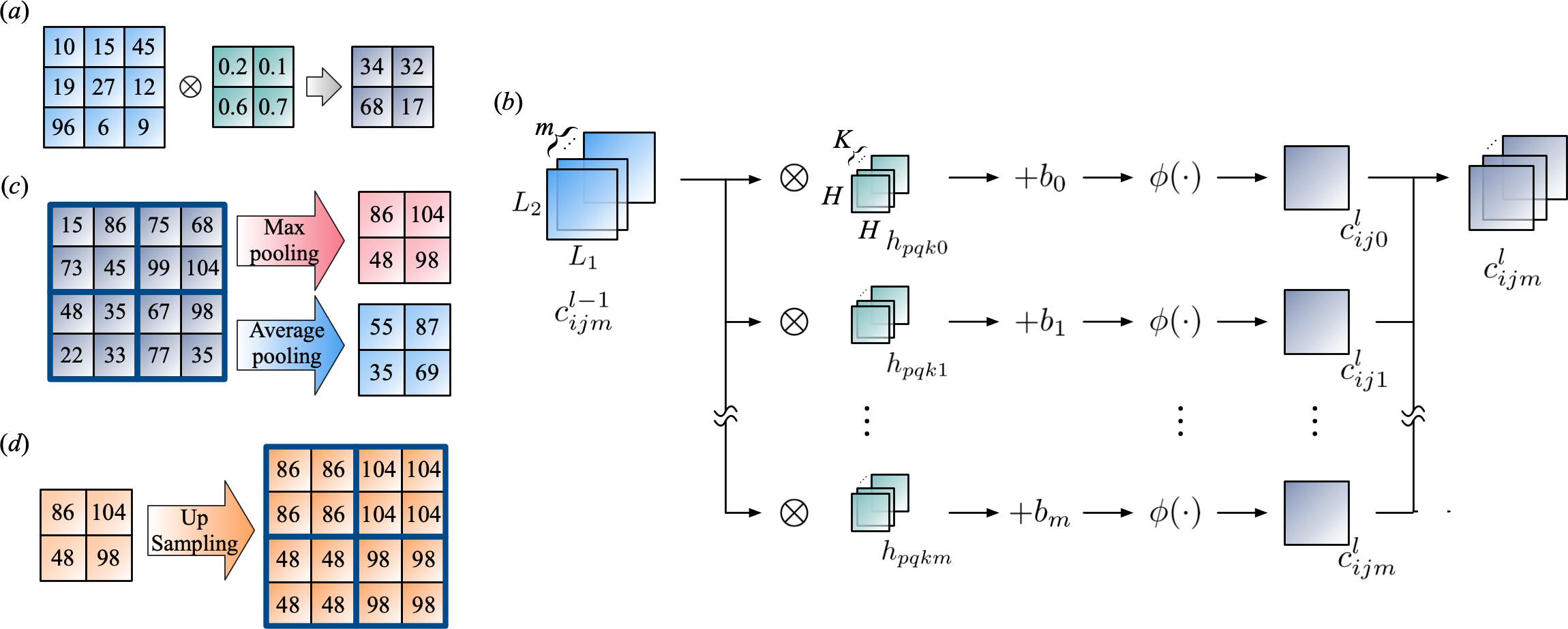}
		\caption{Fundamental operations in convolutional neural network
		(CNN). $(a)$ Convolutional operation with a filter. $(b)$ Computational procedure in the convolution layer. $(c)$ Max and average pooling operations. $(d)$ Upsampling operation.}
		\label{fig_filter}
\end{figure}
%%%%%%%%%%%%%%%%%%%%%%%%%%%%%%%%%%%%%%%%%%%%

We first utilize a convolutional neural network (CNN)~\cite{LBBH1998} for reducing the original dimension of fluid flows.
Nowadays, CNNs have become the essential tools in image recognition thanks to their efficient filtering operation.
Recently, the use of CNNs has also been seen in the community of fluid dynamics~\cite{FukamiVoronoi,LY2019,KL2020,LY2019b,2021wallmodel,nakamura2021comparison}.
A CNN mainly consists of convolutional layers and pooling layers.
As shown in figures~\ref{fig_filter}$(a)$ and $(b)$, the convolutional layer extracts the spatial feature of the input data by a filter operation,
\begin{equation}
    c_{ijm}^{(l)}=\phi\left(\sum^{K-1}_{k=0}\sum^{H-1}_{p=0}\sum^{H-1}_{q=0}c_{i+p-C,j+q-C,k}^{(l-1)}h_{pqkm}+b_{m}\right),
\end{equation}
where $C={\rm floor}(H/2)$, $c_{ijm}^{(l-1)}$ and $c_{ijm}^{(l)}$ are the input and output data at layer $l$, and $b_m$ is a bias.
A filter is expressed as $h_{pqkm}$ with the size of $\left(H\times H\times K\right)$.
In the present paper, the size $H$ for the baseline model is set to 3, although it will be changed for the study on the dependence on the number of weights in section~\ref{sec:numweight}.
The output from the filtering operation is passed through an activation function $\phi$.
In the pooling layer illustrated in figure~\ref{fig_filter}$(c)$, representative values, e.g. maximum value or average value, are extracted from arbitrary regions by pooling operations, such that the image would be downsampled.
It is widely known that the model acquires robustness against the input data by pooling operations thanks to the decrease in the spatial sensitivity of the model~\cite{LNCCK2010}.
Depending on users' task, the upsampling layer can also be utilized.
The upsampling layer copies the value of the low-dimensional images into a high-dimensional space as shown in figure~\ref{fig_filter}$(d)$.
In the training process, the filters in the CNN $h$ are optimized by minimizing a loss function ${E}$ with the back propagation \cite{KB2014} such that ${\bm w}={\rm argmin}_{\bm w}[{E}({\bm q}_{\rm output},{\mathcal F}({\bm q}_{\rm input};{\bm w}))]$.

%%%%%%%%%%%%%%%%%%%%%%%%%%%%%%%%%%%%%%%%%%%%
\begin{figure}
	\vspace{0mm}
	\centering
		%\hspace{-30mm}
		\includegraphics[width=0.80\textwidth]{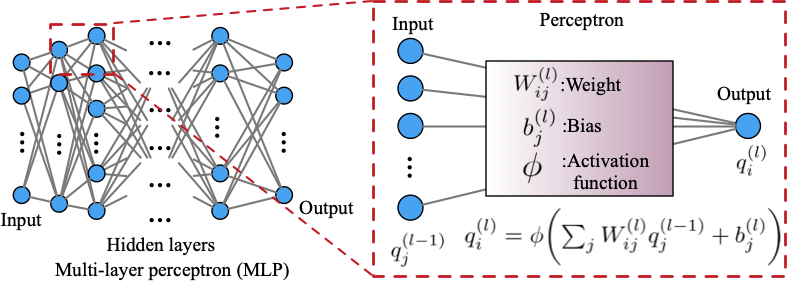}
		\caption{Multi-layer perceptron.}
		\label{fig_mlp}
\end{figure}
%%%%%%%%%%%%%%%%%%%%%%%%%%%%%%%%%%%%%%%%%%%%

We then apply an MLP~\cite{RHW1986} for additional order reduction.
The MLP has been utilized to a wide range of problems, e.g., classification, regression, and speech recognition~\cite{Domingos2012}.  
We can also see the applications of MLP to various problems of fluid dynamics such as turbulence modeling~\cite{ling2016reynolds,gamahara2017searching,MSRV2019}, reduced order modeling~\cite{LW2019}, and field estimation~\cite{YH2019}. 
The MLP can be regarded as the aggregate of a perceptron which is a minimum unit, as illustrated in figure~\ref{fig_mlp}.
The input data from the $(l-1){\rm th}$ layer are multiplied by a weight $\bm{W}$. 
These inputs construct a linear superposition $\bm{Wq}$ with biases $\bm b$ and then passed through a nonlinear activation function $\phi$,
%%Equation1%%
\begin{eqnarray}
    q_i^{(l)}=\phi\biggl(\sum_j W^{(l)}_{ij}q_j^{(l-1)}+b_j^{(l)}\biggr).
    \label{eq:1}
\end{eqnarray}
%%%%%%%%%%%%%%
As illustrated in figure~\ref{fig_mlp}, these perceptrons of MLP are connected with each other and have a fully-connected structure.
As well as the CNN, the weights among all connections $W_{ij}$ are optimized with the minimization manner.

\begin{table}
\begin{center}
  \caption{The present AE structure for the periodic shedding example at the number of latent space $n_r=2$.}
  \label{tab1}
  \small
  \begin{tabular}{ c c | c c} \hline
  \multicolumn{2}{c|}{Encoder} & \multicolumn{2}{c}{Decoder}\\\hline
  Layer&Output shape&Layer&Output shape\\ \hline\hline
  Input&(384, 192, 1)&10th MLP&(4)\\ \hline
  1st Conv2D &(384, 192, 16)&11th MLP&(8)\\ \hline
  1st max pooling &(192, 96, 16)&12th MLP&(16) \\ \hline
  2nd Conv2D &(192, 96, 8)&13th MLP&(32)\\ \hline
  2nd max pooling&(96, 48, 8)&14th MLP&(64)\\ \hline
  3rd Conv2D &(96, 48, 8) &15th MLP&(128)\\ \hline
  3rd max pooling &(48, 24, 8)&16th MLP&(256) \\ \hline
  4th Conv2D &(48, 24, 2)&17th MLP&(512) \\ \hline
  4th max pooling&(24, 12, 2)&18th MLP&(576)\\ \hline
  Flatten &(576)&Reshape&(24, 12, 2)\\ \hline
  1st MLP &(512)&1st upsampling&(48, 24, 2) \\ \hline
  2nd MLP&(256)&5th Conv2D&(48, 24, 2) \\ \hline
  3rd MLP&(128)&2nd upsampling&(96, 48, 2)\\ \hline
  4th MLP&(64)&6th Conv2D&(96, 48, 8)\\ \hline
  5th MLP&(32)&3rd upsampling&(192, 96, 8) \\ \hline
  6th MLP&(16)&7th Conv2D&(192, 96, 8) \\ \hline
  7th MLP&(8)&4th upsampling&(384, 192, 8)\\ \hline
  8th MLP&(4)&8th Conv2D&(384, 192, 16)\\ \hline
  9th MLP (latent space)&(2)&Output & (384, 192, 1) \\ \hline
      \end{tabular}
      \end{center}
\end{table}

\begin{table}
\centering
\caption{{Hyper parameters used in the present AE model.}}
  \begin{tabular}{cccc}
  \hline\hline\noalign{\smallskip}
    Parameter & Value & Parameter & Value \\
    \noalign{\smallskip}\hline\noalign{\smallskip}
    Batch size & 64 & Early stopping patience & 50 \\
    Number of epochs & 5000  & Percentage of training data & 70\% \\
    Learning rate of Adam & 0.001 & Learning rate decay of Adam & 0  \\
    $\beta_1$ of Adam & 0.9 & $\beta_2$ of Adam & 0.999 \\
    \noalign{\smallskip}\hline\hline
  \end{tabular}
  \label{tab:param}
\end{table}

By combining two methods introduced above, we here construct the AE as illustrated in figure~\ref{fig_ae}.
The AE has a lower dimensional space called the latent space at the bottleneck so that a representative feature of high-dimensional data $\bm q$ can be extracted by setting the target data as both the input and output.
In other words, if we can obtain the similar output to the input data, the dimensional reduction can be successfully achieved onto the latent vector $\bm r$.
In turn, with over-compression of the original data $\bm q$, of course, the decoded field $\widehat{\bm q}$ cannot be remapped well.
In sum, these relations can be formulated as
\begin{align}
    {\bm{r}} ={\mathcal F}_{e}({\bm q}), ~~{\bm q}\approx\widehat{\bm{q}} ={\mathcal F}_{d}(\bm r),
\end{align}
where ${\cal F}_e$ and ${\cal F}_d$ are encoder and decoder of autoencoder as shown in figure \ref{fig_ae}.
Mathematically, in the training process for obtaining the autoencoder model $\cal F$, the weights $\bm w$ is optimized to minimize an error function $E$ such that ${\bm w}={\rm argmin}_{\bm w}[{E}({\bm q},{\mathcal F}({\bm q};{\bm w}))]$.

In the present study, we use the $L_2$ norm error as the cost function $E$ and Adam optimizer~\cite{KB2014} is applied for obtaining the weights.
Other parameters used for construction of the present AE are summarized in table~\ref{tab:param}.
{Note that theoretical optimization methods~\cite{BYC2013,BCF2009,maulik2019time} can be considered for hyperparameter decision to further improve the AE capability, although not used in the present study.}
To avoid an overfitting, we set the early stopping criteria~\cite{prechelt1998}.
With regard to the internal steps of the present AE, we first apply the CNN to reduce the spatial dimension of flow fields by ${\cal O}(10^2)$.
The MLP is then used to obtain the latent modes by feeding the compressed vector by the CNN, as summarized in table~\ref{tab1}, which shows an example of the AE structure for the cylinder wake problem with the number of latent variables $n_r=2$.
For a case where it is considered irrational to reduce the number of latent variables down to ${\cal O}(10^2)$, e.g., $n_r=3072$ of turbulent channel flow, the order reduction is conducted by means of CNN only (without MLP).

\subsection{The role of activation functions in autoencoder}
\label{sec:activation_intro}

One of key features in the mathematical procedure of AE is the use of activation functions.
This is exactly the reason why neural networks can account for nonlinearlity into their estimations.
Here, let us mathematically introduce the strength of nonlinear activation functions for model order reduction by comparing to a proper orthogonal decomposition (POD), which has been known as equivalent to AE with the linear activation function $z={\phi}(z)$.
The details can be seen in some previous studies
\cite{BH1989,bourlard1988auto,oja1982simplified}.

%%%%%%%%%%%%%%%%%%%%%%%%%%%%%%%%%%%%%%%%%%%%
\begin{figure}
    \centering
    \includegraphics[width=0.65\textwidth]{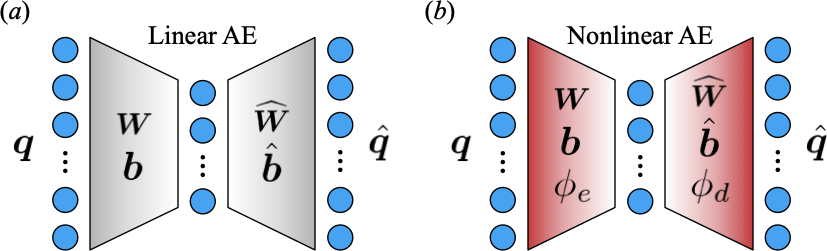}
    \caption{$(a)$ Linear autoencoder. $(b)$ Autoencoder with nonlinear activation functions $\phi_e$ and $\phi_d$.}
    \label{fig:AEPOD}
\end{figure}
%%%%%%%%%%%%%%%%%%%%%%%%%%%%%%%%%%%%%%%%%%%%

We first visit the POD-based order reduction for high-dimensional data matrix ${\bm q}\in {\mathbb R}^{n_{t}\times n_{g}}$ with a reduced basis ${\bm\Gamma}\in {\mathbb R}^{n_{r}\times n_{g}}$, where $n_t$, $n_r$ and $n_g$ denote the numbers of collected snapshots, latent modes and spatial discretized grids, respectively.
An optimization procedure of POD can be formulated with an $L_2$ minimization manner,
\begin{equation}\label{eq:pod}
  {\bm\Gamma}={\rm argmin}_{\bm\Gamma}||(\bm q-\overline{\bm q})-{\bm\Gamma}^{\rm T}{\bm\Gamma}(\bm q-\overline{\bm q})||_2, 
\end{equation}
where $\overline{\bm q}$ is the mean value of the high-dimensional data matrix.
Next, we consider an AE with the linear activation function $z={\phi}(z)$, as shown in figure \ref{fig:AEPOD}$(a)$.
Analogous to equation \ref{eq:pod}, the optimization procedure can be written as 
\begin{eqnarray}
{{\bm \omega}}={\rm argmin}_{\bm w}||{\bm q}-\widehat{\bm q}||_2,~~~ {\rm where}~~~\widehat{\bm q}=\widehat{\bm W}({\bm W}{\bm q}+{\bm b})+\widehat{\bm b},
\end{eqnarray}
with a reconstructed matrix $\widehat{\bm q}$, and weights ${\bm w}=[{\bm W},{\bm b},\widehat{\bm W},\widehat{\bm b}]$ in the AE,
 which contains the weights of encoder $\bm W$ and decoder $\widehat{\bm W}$ and the biases of encoder $\bm b$ and decoder $\widehat{\bm b}$.
Substituting these variables with ${\bm q}$ for equation \ref{eq:pod}, the {argmin} formulation can be transformed as 
\begin{align}\label{eq:ae}
{{\bm \omega}}&={\rm argmin}_{\bm w}||{\bm q}-\widehat{\bm q}||_2\nonumber\\
&= {\rm argmin}_{\bm w} ||{\bm q}-{\widehat {\bm W}({\bm W}{\bm q}+{\bm b})-{\widehat{\bm b}}}||_2\nonumber\\
&={\rm argmin}_{\bm w}||({\bm q}-\widehat{\bm b})-(\widehat{\bm W} {\bm W}{\bm q}+\widehat{\bm W}{\bm b})||_2.
\end{align}
Noteworthy here is that equations \ref{eq:pod} and \ref{eq:ae} are equivalent with each other by replacing the variables as follows,
\begin{eqnarray}
{\bm W}\rightarrow{\bm \Gamma},~~\widehat{\bm W}\rightarrow{\bm \Gamma}^{\rm T},~~\widehat{\bm b}\rightarrow \overline{\bm q},~~ {\bm b}\rightarrow{-\overline{\bm q}{\bm\Gamma}}.
\end{eqnarray}
These transformations tell us that the AE with the linear activation is essentially the same order reduction tool as POD.
In other words, with regard to the AE with nonlinear activation functions as illustrated in figure \ref{fig:AEPOD}$(b)$, the nonlinear activation functions of encoder $\phi_e$ and decoder $\phi_d$ are multiplied to the weights (reduced basis) in equation \ref{eq:pod}, such that
\begin{equation}
  {\bm\Gamma}={\rm argmin}_{\bm\Gamma}||(\bm q-\overline{\bm q})-\phi_d{\bm\Gamma}^{\rm T}\phi_e{\bm\Gamma}(\bm q-\overline{\bm q})||_2.
\end{equation}
Note that the weights $\bm w$ including biases are optimized in the training procedure of autoencoder. 

\begin{figure}
    \centering
    \includegraphics[width=1.00\textwidth]{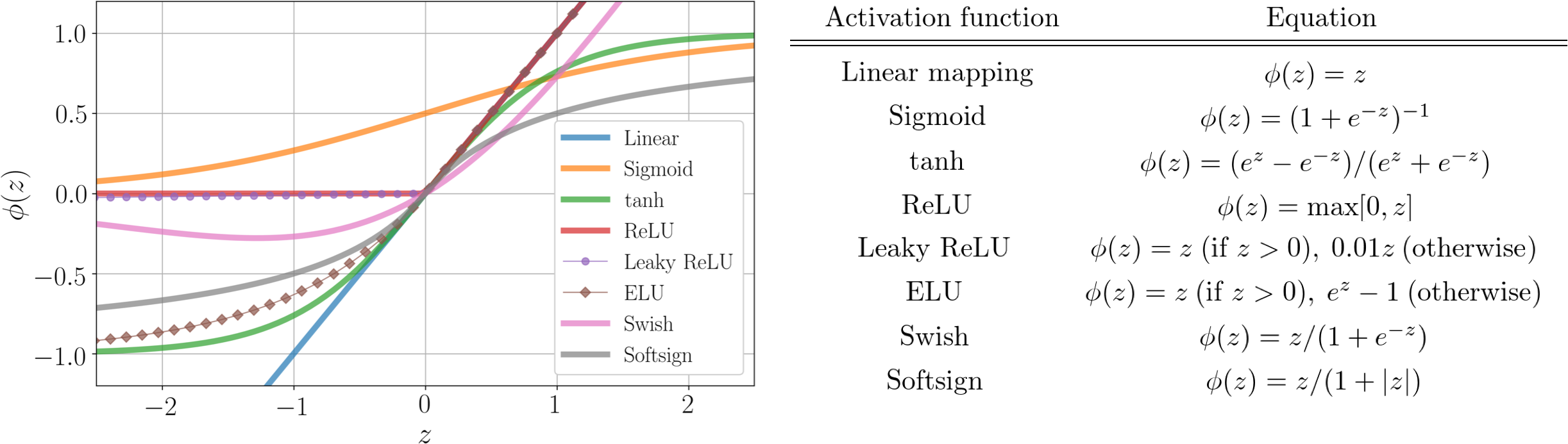}
    \caption{Activation functions covered in the present study.}
    \label{fig:act}
\end{figure}

As discussed above, the choice of nonlinear activation function is crucial to outperform the traditional linear method.
In this study, we consider the use of various activation functions as summarized in figure~\ref{fig:act}.
As shown, we cover a wide nature of nonlinear functions which are the well-used candidates, although these are not all of 
the used functions in the machine learning field.
We refer the readers to Nwankpa et al.~\cite{nwankpa2018activation} for details of each nonlinear activation function.
For the baseline model of each example, we use the Rectified Linear Unit (ReLU)~\cite{NH2010}, which has been widely known as a good candidate in terms of weight updates. 
Note in passing that we use the same activation function at all layers in each model; namely, a case like ReLU for the 1st and 2nd layers and tanh for the 2nd and 3rd layers is not considered.
In the present paper, we perform a three-fold cross validation~\cite{Bruntonkutz2019}, although the representative flow fields with ensemble error values are reported.
{For readers' information, the training cost with the representative number of latent variables for each case performed under the NVIDIA TESLA V100 graphics processing unit (GPU) is summarized in table~\ref{tab_cost}.

\begin{table}
\centering
  \caption{{Training time of AE in the present study.}}
\begin{tabular}{ccc}
\hline\hline
Examples & Number of latent variables $n_{r}$ & Training time \\\hline
Cylinder wake  & 128   & 1433.75 s $=$ 1.55 s/epoch $\times$ 925 epochs    \\
Transient  & 128   & 31665.20 s $=$ 30.1 s/epoch $\times$ 1052 epochs      \\
Sea surface temperature  & 128   & 5883.82 s $=$ 2.18 s/epoch $\times$ 2699 epochs     \\
Turbulent channel flow  & 384   & 1342.60 s $=$ 0.70 s/epoch $\times$ 1918 epochs \\
\hline\hline
\end{tabular}
\label{tab_cost}
\end{table}
}

\section{Setups for covered examples of fluid flows}
\label{sec:data}

%%%%%%%%%%%%%%%%%%%%%%%%%%%%%%%%%%%%%%%%%%%%
\begin{figure}
    \centering
    \includegraphics[width=1.00\textwidth]{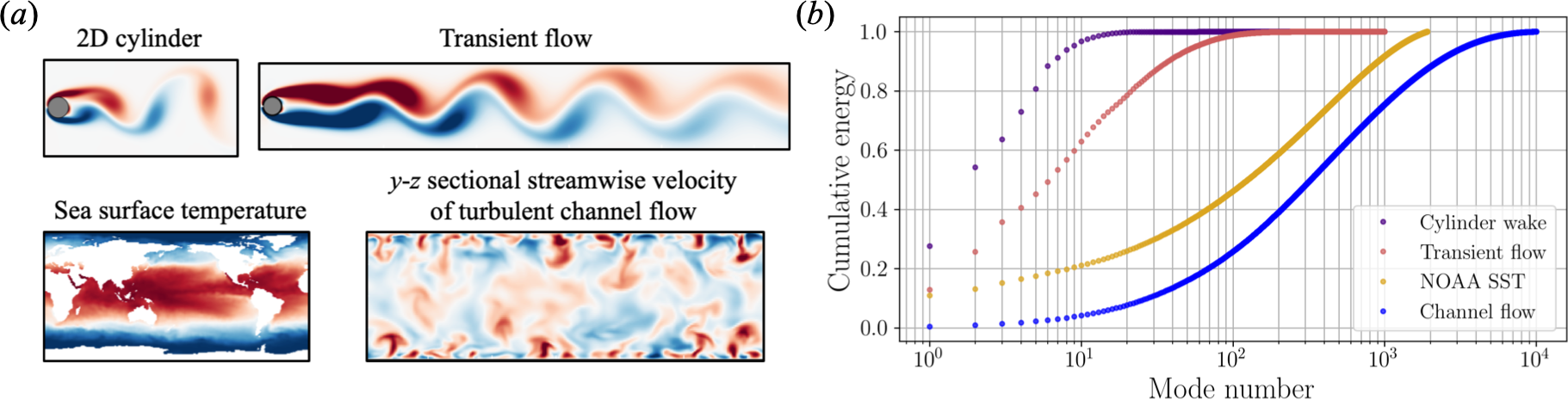}
    \caption{$(a)$ Examples of flow fields used in the present study. $(b)$ Normalized cumulative sum of the singular values.}
    \label{fig_flows}
\end{figure}
%%%%%%%%%%%%%%%%%%%%%%%%%%%%%%%%%%%%%%%%%%%%

In the present study, we cover a wide range of fluid flow nature considering four examples which include laminar and turbulent flows, as shown in figure~\ref{fig_flows}$(a)$.
The broad spread of complexity in the covered example flows can be also seen in the cumulative singular value spectra presented in figure~\ref{fig_flows}$(b)$.

\subsection{Two-dimensional cylinder wake at $Re_D=100$}

The first example is a two-dimensional cylinder wake at ${Re}_D=100$.  
The data set is prepared using a two-dimensional direct numerical simulation (DNS)~\cite{kor2017}. 
The governing equations are the continuity and the incompressible Navier--Stokes equation,
\begin{align}
    \bm{\nabla} \cdot \bm{u}=0,~~~\dfrac{\partial\bm{u}}{\partial t} + \bm{\nabla} \cdot (\bm{uu})  = - \bm{\nabla} p + \dfrac{1}{{Re}_D}\nabla ^2 \bm{u},
\end{align}
where $\bm{u}=[u,v]$ and $p$ are the velocity vector and pressure, respectively.
All quantities are non-dimensionalized with the fluid density, the free-stream velocity, and the cylinder diameter.
The size of the computational domain is ($L_x, L_y$)=(25.6, 20.0),
and the cylinder center is located at $(x, y)=(9,0)$.
A Cartesian grid with the grid spacing of $\Delta x=\Delta y = 0.025$ is used for the numerical setup, and the number of grid points for the present DNS is $(N_x, N_y)=(1024, 800)$.  
For the machine learning, only the flow field around the cylinder is extracted as the data set to be used such that $8.2 \leq x \leq 17.8$ and $-2.4 \leq y \leq 2.4$ with $(N_x^*, N_y^*)=(384, 192)$.  
We use the vorticity field $\omega$ as the data attribute. 
The time interval of flow data is 0.25 corresponding to approximately 23 snapshots per a period with the Strouhal number equals to 0.172.
For training the present AE model, we use 1000 snapshots.

\subsection{Transient wake at $Re_D=100$}

We then consider a transient wake behind the cylinder, which is not periodic in time. 
Since almost the same computational procedure is applied as the case of periodic shedding, let us briefly touch the notable contents.
The training data is obtained from the DNS similarly to the first example.
Here, the domain is extended 3 folds in $x$ direction such that $(N^*_x,N^*_y)=(1152,192)$ with $(L^*_x,L^*_y)=(28.8,4.8)$.
Also similarly to the first example, we use the vorticity field $\omega$ as the data attribute.
For training the present AE model, we use 4000 snapshots whose sampling interval is finer than that in the periodic shedding case to capture the transient nature correctly.

\subsection{NOAA sea surface temperature}

The third example is the NOAA sea surface temperature data set~\cite{noaa} obtained from satellite and ship observations.
The aim in using this data set is to examine the applicability of AE to more practical situations, such as the case without governing equations.
We should note that this data set is driven by the influence of seasonal periodicity.
The spatial resolution here is $360\times180$ based on a one degree grid.
We use the 20 years of data corresponding to 1040 snapshots spanning from year 1981 to 2001.  

\subsection{$y-z$ sectional field of turbulent channel flow at $Re_{\tau}=180$}

To investigate the capability of AE for chaotic turbulence, we lastly consider the use of a cross-sectional ($y-z$) streamwise velocity field of a turbulent channel flow at ${Re}_{\tau}=180$ under a constant pressure gradient condition.
The data set is prepared by the DNS code developed by Fukagata et al.~\cite{FKK2006}.
The governing equations are the incompressible Navier--Stokes equations, i.e.,
\begin{eqnarray} 
\bm{\nabla} \cdot {\bm u} = 0,~~~
{ \dfrac{\partial {\bm u}}{\partial t}  + \bm{\nabla} \cdot ({\bm u \bm u}) =  -\bm{\nabla} p  + \dfrac{1}{{Re}_\tau}\nabla^2 {\bm u}},
\end{eqnarray}
where $\displaystyle{{\bm u} = [u~v~w]^{\mathrm T}}$ represents the velocity with $u$, $v$ and $w$ in the streamwise ($x$), the wall-normal ($y$) and the spanwise ($z$) directions.  
Also, $t$ is the time, $p$ is the pressure, and ${{Re}_\tau = u_\tau  \delta/\nu}$ is the friction Reynolds number. 
The quantities are non-dimensionalized with the channel half-width $\delta$ and the friction velocity $u_\tau$.
The size of the computational domain and the number of grid points are $(L_{x}, L_{y}, L_{z}) = (4\pi\delta, 2\delta, 2\pi\delta)$ and $(N_{x}, N_{y}, N_{z}) = (256, 96, 256)$, respectively.  
The grids in the $x$ and $z$ directions are uniform, and a non-uniform grid is applied in the $y$ direction.  
The no-slip boundary condition is applied to the walls and the periodic boundary condition is used in the $x$ and $z$ directions.  
We use the fluctuation component of a $y-z$ sectional streamwise velocity $u^\prime$ as the representative data attribute for the present AE.

\section{Results}
\label{sec:result}

In this section, we examine the influence on the number of latent modes (sec.~\ref{sec:num_lat}), the choice of activation function for hidden layers (sec.~\ref{sec:caf}), and the number of weights contained in an autoencoder (sec. \ref{sec:numweight}) with the data sets introduced above.

\subsection{Number of latent modes}
\label{sec:num_lat}

Let us first investigate the dependence of reconstruction accuracy by autoencoder (AE) on the number of latent variables
$n_r$.
The expected trend in this assessment for all data sets is that the reconstruction error would increase with reducing $n_r$, since the AE needs to discard the information while extracting dominant features from high-dimensional flow fields into a limited dimension of latent space.

%%%%%%%%%%%%%%%%%%%%%%%%%%%%%%%%%%%%%%%%%%%%
\begin{figure}
    \centering
    \includegraphics[width=0.98\textwidth]{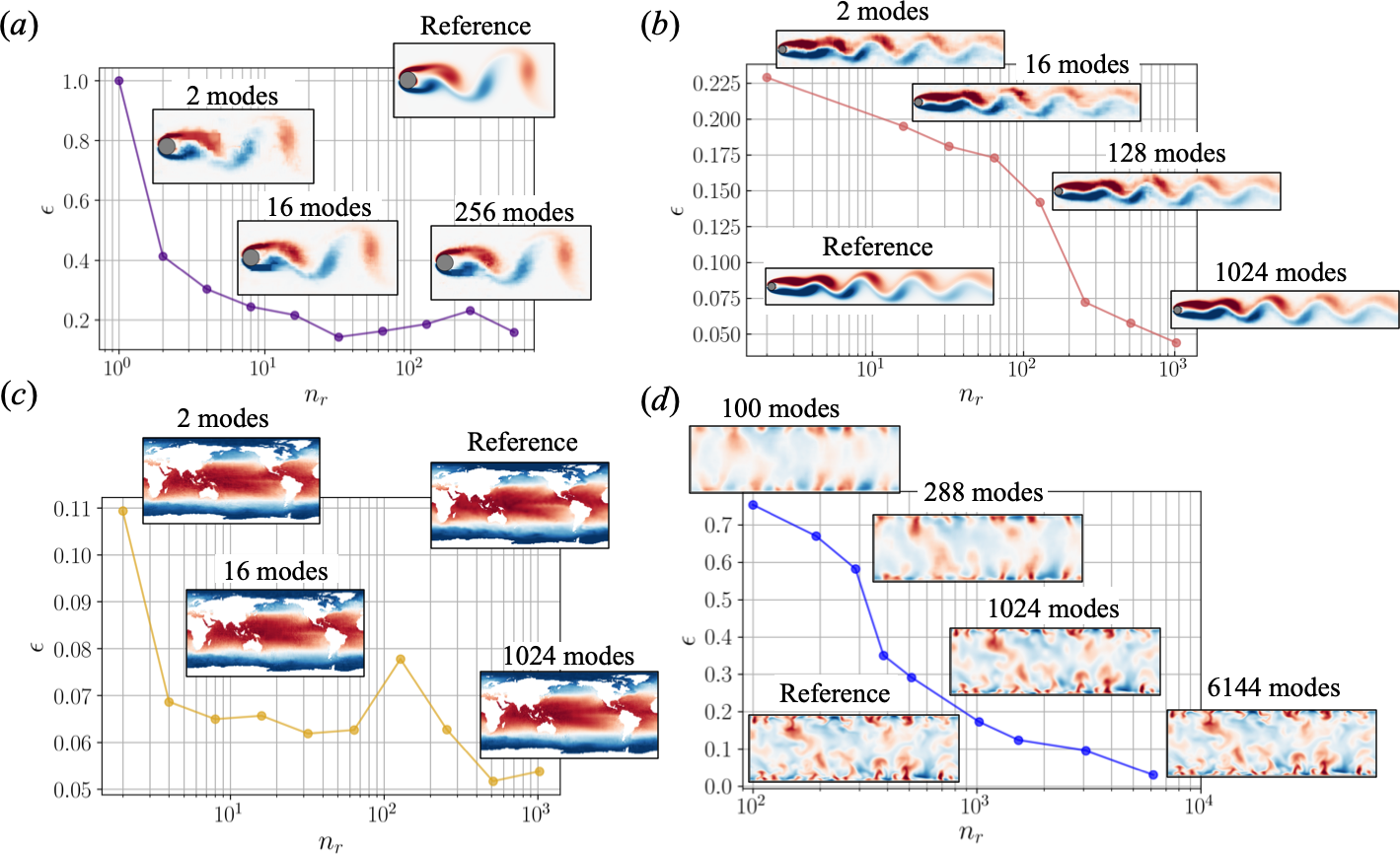}
    \caption{The relationship between the number of latent variables $n_r$ and $L_2$ error norm $\epsilon = ||{\bm q}_{\rm Ref}-{\bm q}_{\rm ML}||_2/||{\bm q}_{\rm Ref}||_2$. $(a)$ Two-dimensional cylinder wake. $(b)$ Transient flow. $(c)$ NOAA sea surface temperature. $(d)$ $y-z$ sectional streamwise velocity fluctuation of turbulent channel flow. Three-fold cross validation is performed, although not shown.}
    \label{fig_nlm}
\end{figure}
%%%%%%%%%%%%%%%%%%%%%%%%%%%%%%%%%%%%%%%%%%%%

The relationship between the number of latent variables $n_r$ and $L_2$ error norm $\epsilon = ||{\bm q}_{\rm Ref}-{\bm q}_{\rm ML}||_2/||{\bm q}_{\rm Ref}||_2$, where ${\bm q}_{\rm Ref}$ and ${\bm q}_{\rm ML}$ indicate the reference field and the reconstructed field by an AE, is summarized in figure~\ref{fig_nlm}.
For the periodic shedding case, the structure of whole field can be captured with only two modes as presented in figure~\ref{fig_nlm}$(a)$.
However, the $L_2$ error norm at $n_r=2$ is approximately 0.4, which shows a relatively large value despite the fact that the periodic shedding behind a cylinder at $Re_D=100$ can be represented by only one scalar value~\cite{TN2018}.
The error value here is also larger than the reported $L_2$ error norms in the previous studies~\cite{MFF2019,FNF2020} which handle the cylinder wake at $Re_D=100$ using AEs with $n_r=2$.
This is likely due to the choice of activation function --- in fact, we find a significant difference depending on the choice, which will be explained in section~\ref{sec:caf}.
It is also striking that the $L_2$ error at $n_r=10^1-10^2$ somehow increases.
This indicates that the number of weights $n_w$ contained in the AE has also a non-negligible effect for the mapping ability, since the number of weights is increasing with $n_r$ in our AE setting as can be seen in table \ref{tab1}, e.g., $n_w(n_r=128)>n_w(n_r=256)$.
This point will be investigated in section~\ref{sec:numweight}.
We then apply the AE to the transient flow as shown in figure~\ref{fig_nlm}$(b)$.
Since the flow in this transient case is also laminar, the whole trend of wake can be represented with a few number of modes, e.g., 2 and 16 modes.
The behavior of AE with the data which has no modelled governing equation can be seen in figure~\ref{fig_nlm}$(c)$.
As shown, the entire trend of the field can be extracted with only two modes since the considered data here is driven by the seasonal periodicity as mentioned above.
The reason for a rise in the error at $n_r=128$ is likely because of the influence on the number of weights, which is also observed in the cylinder example.
We can see a striking difference of AE performance against the previous three examples in figure~\ref{fig_nlm}$(d)$.
For the turbulent case, the required number of modes to represent finer scales is visibly larger than those in the other cases since turbulence has a much wider range of scales.
The difficulty in compression for turbulence can also be found in terms of $L_2$ error norms.

%%%%%%%%%%%%%%%%%%%%%%%%%%%%%%%%%%%%%%%%%%%%
\begin{figure}
    \centering
    \includegraphics[width=0.98\textwidth]{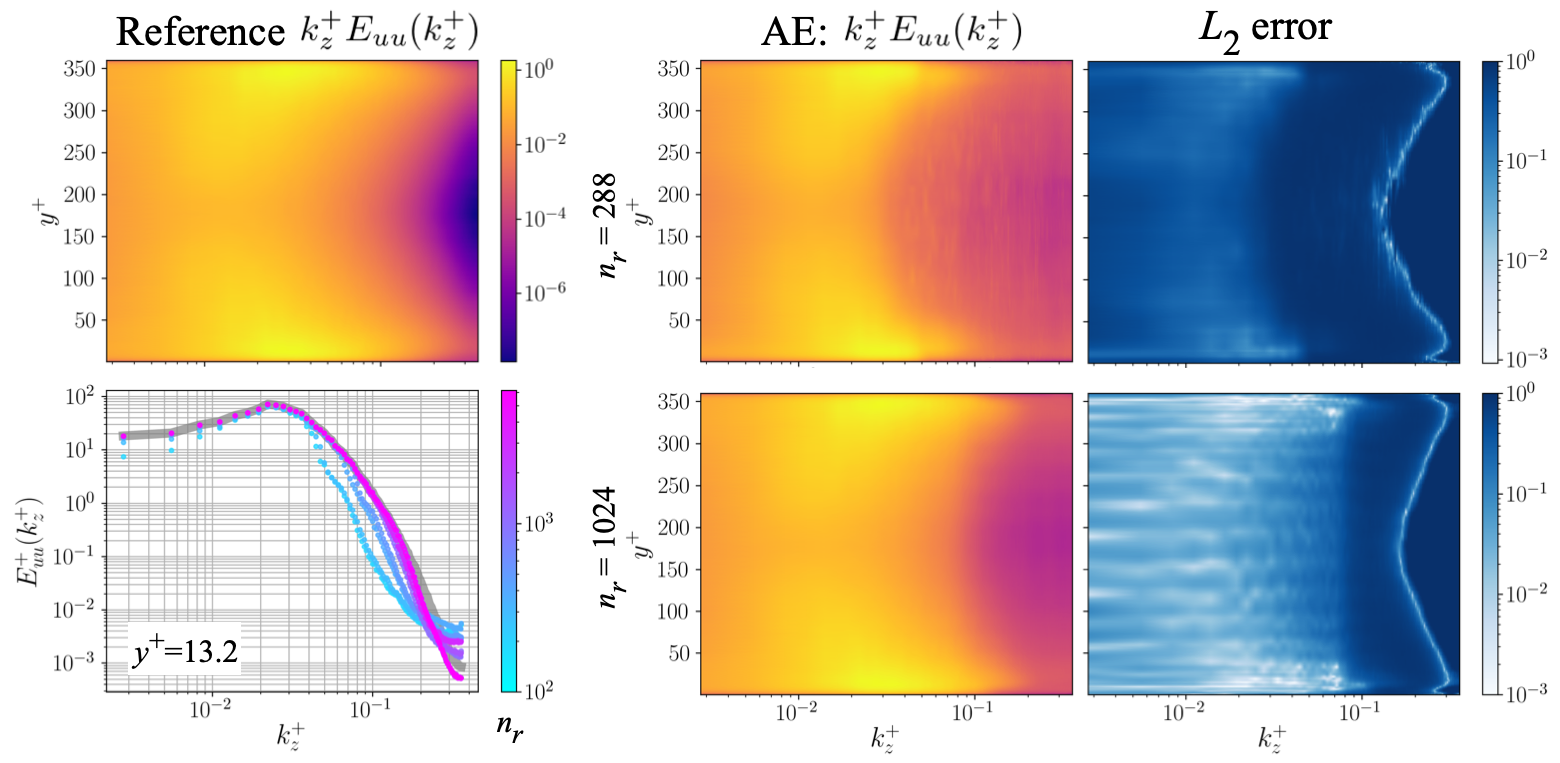}
    \caption{Premultiplied spanwise kinetic energy spectrum at $n_r=288$ and 1024. The lower left shows $E^+_{uu}(k_z^+)$ at $y^+=13.2$; in this plot, the gray line indicates the reference DNS and the other colored plots correspond to the AE models.}
    \label{fig_en}
\end{figure}
%%%%%%%%%%%%%%%%%%%%%%%%%%%%%%%%%%%%%%%%%%%%

To further examine the dependence of reconstruction ability by AEs on the number of latent modes in terms of scales in turbulence, we check the premultiplied spanwise kinetic turbulent energy spectrum $k_z^+E^+_{uu}(k_z^+)$ in figure~\ref{fig_en}. 
As shown here, the error level $||k^+_zE^+_{uu}(k^+_z)_{\rm Ref}-k^+_zE^+_{uu}(k^+_z)_{\rm ML}||_2/||k^+_zE^+_{uu}(k^+_z)_{\rm Ref}||_2$ decreases with increasing $n_r$, especially on the low-wavenumber space.
This implies that the AE model extracts the information in the low-wavenumber region preferentially.
Noteworthy here is that the low error level portion suddenly appears in the high-wavenumber region, i.e., white lines at $n_r=288$ and 1024 in the $L_2$ error map of figure~\ref{fig_en}.
These correspond to the crossing point of the $E^+_{uu}(k^+_z)$ curves of DNS and AE as can be seen in the lower left of figure \ref{fig_en}.
The under- or overestimation of the kinetic turbulent energy spectrum is caused by a combination of several reasons, e.g., underestimation of $u^\prime$ because of $L_2$ fitting and the relationship between a squared velocity and energy such that $\overline{u^2} = \int E^+_{uu}(k^+_z) dk^+_z$, although it is difficult to separate these into each element.
Note that the aforementioned observation is also seen in Scherl et al.~\cite{SSSWPB2020}, who applied a robust principal component analysis to the channel flow.

\subsection{Choice of activation function}
\label{sec:caf}

Next, we examine the influence of the AE based low dimensionalization on the choice of activation functions.
As presented in figure~\ref{fig:act}, we cover a wide range of functions: namely linear mapping, sigmoid, hyperbolic tangent (tanh), rectified linear unit (ReLU, baseline), Leaky ReLU, ELU, Swish, and Softsign.
Since the use of activation function is a essential key to account for nonlinearities into the AE-based order reduction as discussed in section~\ref{sec:activation_intro}, the investigation of the choice here can be regarded as immensely crucial for the construction of AE.
For each fluid flow example, we consider the same AE construction and parameters as the models at the first assessment (except for the activation function), which achieved the $L_2$ error norm of approximately 0.2.
Note that an exception is the example of sea surface temperature, since the highest reported $L_2$ error over the covered number of latent modes is already lower than 0.2.
Hence, the numbers of latent modes for each example here are $n_r=16$ (cylinder wake), 16 (transient), 2 (sea surface temperature), and 1024 (turbulent channel flow), respectively.

%%%%%%%%%%%%%%%%%%%%%%%%%%%%%%%%%%%%%%%%%%%%
\begin{figure}
    \centering
    \includegraphics[width=1.00\textwidth]{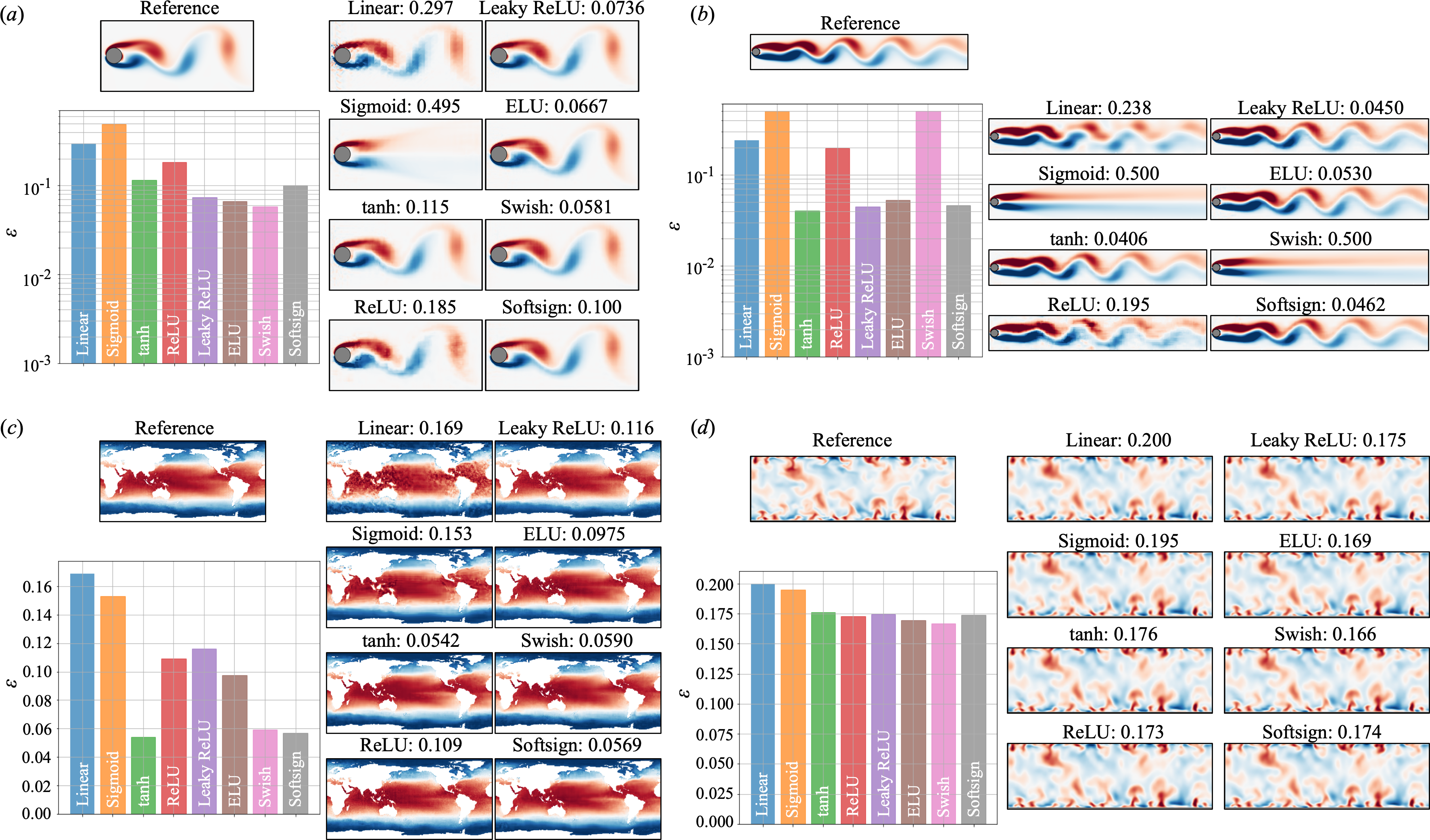}
    \caption{Dependence of mapping ability on activation functions. $(a)$ Two-dimensional cylinder wake. $(b)$ Transient flow. $(c)$ NOAA sea surface temperature. $(d)$ $y-z$ sectional streamwise velocity fluctuation of turbulent channel flow. Three-fold cross validation is performed, although not shown.  Values above contours indicate the $L_2$ error norm.}
    \label{fig_caf}
\end{figure}
%%%%%%%%%%%%%%%%%%%%%%%%%%%%%%%%%%%%%%%%%%%%

The dependence of the AE reconstruction on activation functions is summarized in figure \ref{fig_caf}.
As compared to the linear mapping $\phi(z)=z$, the use of nonlinear activation functions leads to improve the accuracy with almost all cases.
However, as shown in figure~\ref{fig_caf}$(a)$, the field 
reconstructed with sigmoid function $\phi(z)=(1+e^{-z})^{-1}$ is worse than the others including linear mapping, which shows the mode 0-like  field.
This is likely because of the widely known fact that the use of sigmoid function has often encountered the vanishing gradient which cannot update weights of neural networks accordingly~\cite{nwankpa2018activation}.
The same trend can also be found with the transient flow, as presented in figure~\ref{fig_caf}$(b)$.
To overcome the aforementioned issue, the swish function $\phi(z)=z/(1+e^{-z})$, which has a similar form to sigmoid, was proposed~\cite{RZL2017}.
This swish can improve the accuracy of the AE based order reduction with the example of periodic shedding, although it is not for the transient.
These observations indicate the importance to check the ability of each activation function so as to represent an efficient order reduction.
The other notable trend here is the difference of the influence on the selected activation functions depending on the target flows.
In our investigation, the appropriate choice of nonlinear activation functions exhibits significant improvement for the first three problem settings, i.e., periodic wake, transient, and sea surface temperature.
In contrast, we do not see substantial differences among the nonlinear functions with the turbulence example.
It implies that a range of scales contained in flows has also a great impact on the mapping ability of AE with nonlinear activation functions.
Summarizing above, a careful choice of activation functions should be taken to implement the AE with acceptable order reduction since the influence and abilities of activation functions highly depend on the target flows which users deal with.

\subsection{Number of weights}
\label{sec:numweight}

As discussed in section \ref{sec:num_lat}, the number of weights contained in AEs may also have an influence for the mapping ability.
Here, let us examine this point with two investigations as follows:
\begin{enumerate}
    \item by changing the amount of parameters in the present AE. (section \ref{sec:cpp})
    \item by pruning weights in the present AE. (section \ref{sec:pru})
\end{enumerate}

\subsubsection{Change of parameters}
\label{sec:cpp}

As mentioned above, the present AE comprises of a convolutional neural network (CNN) and multi-layer perceptrons (MLP).
For the assessment here, to change the number of weights in the AE, we tune various parameters in both the CNN and the MLP, e.g., the number of hidden layers in both CNN and MLP, the number of units in MLP, the size and number of filters in CNN.
Note in passing that we here only focus on the number of weights, although the way to align the number of weights may also have effects to AE's ability.
For example, we do not consider difference between two cases of $n_{w,1}=n_{w,2}$ --- one model with $n_{w,1}$ is obtained by tuning the number of layers, while the other with $n_{w,2}$ is constructed by changing the number of units --- although the performance of two may be different caused by the tuned parameters.
The same AE construction and parameters as those used in the second assessment with ReLU function are considered, which achieved the $L_2$ error norm of approximately 0.2 in the first investigation.
Similarly to the investigation for activation functions, the numbers of latent modes $n_r$ for each example are 16 (cylinder wake), 16 (transient), 2 (sea surface temperature), and 1024 (turbulent channel flow), respectively.
The dependence of the mapping ability on $n_w$ is investigated as follows,
\begin{enumerate}
    \item The number of weights in the MLP $n_{w,{\rm MLP}}$ is fixed,
     and the number of weights in the CNN $n_{w,{\rm CNN}}$ is changed. The $L_2$ error is assessed 
     by
     comparing the original number of weights in the CNN such that $n_{w,{\rm CNN}}/n_{w,{\rm CNN:original}}$.
    \item The number of weights in the CNN $n_{w,{\rm CNN}}$ is fixed,
     and the number of weights in the MLP $n_{w,{\rm MLP}}$ is changed. The $L_2$ error is assessed 
     by
     comparing the original number of weights in the MLP such that $n_{w,{\rm MLP}}/n_{w,{\rm MLP:original}}$.
\end{enumerate}

\begin{table}
\centering
  \caption{The number of weights in the baseline AEs for section~\ref{sec:cpp}.}
\begin{tabular}{cccc}
\hline\hline
Examples & All $n_{w}$ & CNN $n_{w,{\rm CNN}}$ & MLP $n_{w,{\rm MLP}}$ \\\hline
Cylinder wake  & 945721   & 4137   & 941584   \\
Transient  & 270057   & 5305   & 264752   \\
Sea surface temperature  & 59311   & 37093   & 22218   \\
Turbulent channel flow  & 21930   & 21930   & --- \\
\hline\hline
\end{tabular}
\label{tab2}
\end{table}

The number of weights $n_w=n_{w,{\rm CNN}}+n_{w,{\rm MLP}}$ in the baseline models is summarized in table~\ref{tab2}.
Note again that the CNN first works to reduce the dimension of flows by ${\cal{O}}(10^2)$ and the MLP is then utilized to map into the target dimension as stated in section~\ref{sec:AEscheme}.
Hence, the considered models for the cylinder wake, transient flow, and sea surface temperature have both the CNN and the MLP.
On the other hand, the MLP is not applied to the model for the example of turbulent channel flow, since $n_r=1024$.
Following this reason, only the influence on the weights of CNN is considered for the turbulent case.

%%%%%%%%%%%%%%%%%%%%%%%%%%%%%%%%%%%%%%%%%%%%
\begin{figure}
    \centering
    \includegraphics[width=0.75\textwidth]{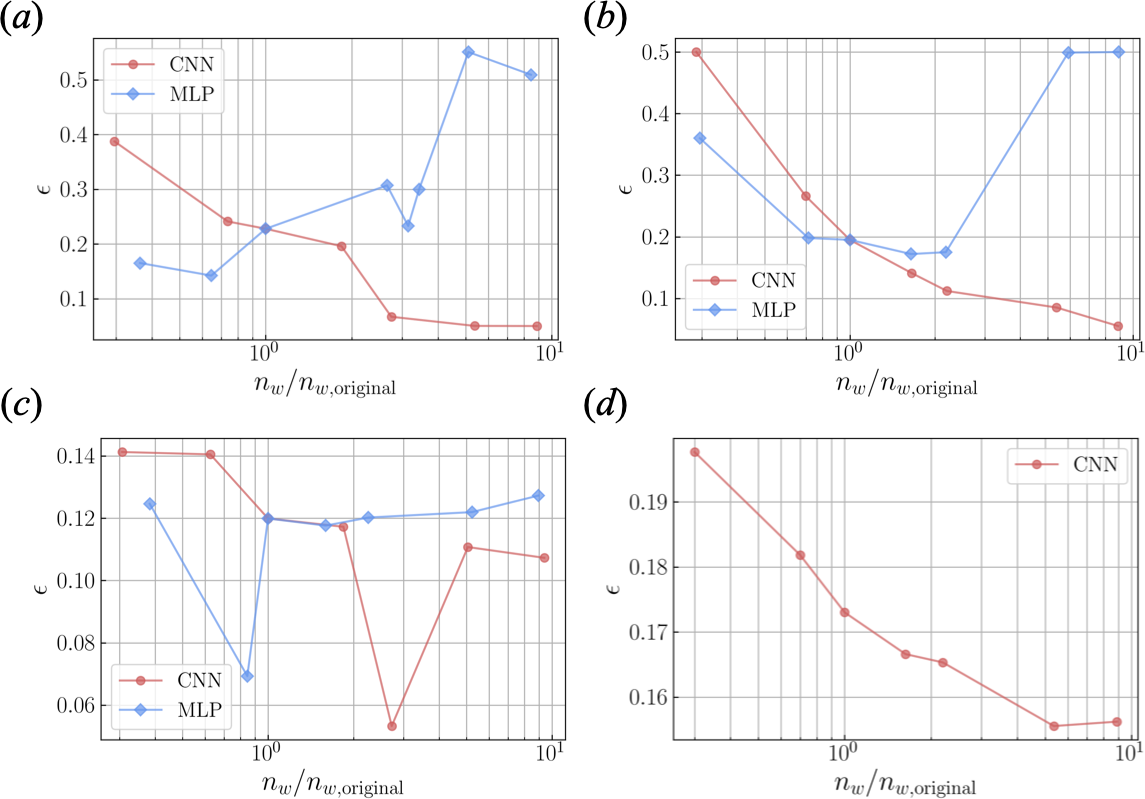}
    \caption{Dependence on the number of weights tuned by modification of parameters in AE.  $(a)$ Two-dimensional cylinder wake. $(b)$ Transient flow. $(c)$ NOAA sea surface temperature. $(d)$ $y-z$ sectional streamwise velocity fluctuation of turbulent channel flow.}
    \label{fig_nw1}
\end{figure}
%%%%%%%%%%%%%%%%%%%%%%%%%%%%%%%%%%%%%%%%%%%%

The dependence on the number of weights tuned by changing of parameters is summarized in figure~\ref{fig_nw1}.
The expected trend here is that the error decreases with increasing the number of weights, which is seen with the example of turbulent channel flow shown in figure~\ref{fig_nw1}$(d)$.
The same trend can also be observed with the parameter modification in CNN for the examples of periodic shedding and transient cylinder flows, as presented in figures~\ref{fig_nw1}$(a)$ and $(b)$.
The error curve for the parameter modification in MLP, however, shows the contrary trend in both cases --- the error increases for the larger number of weights.
This is likely because there are too many numbers of weights in MLP, and this deepness leads to a
difficulty in weight updates, even if we use the ReLU function.
With regard to the models of sea surface temperature, the error curves of both CNN and MLP show a valley-like behavior.
It implies that these numbers of weights which achieve the lowest errors are appropriate in terms of weight updates.
For the channel flow example, the error converges around 0.15 as shown in figure~\ref{fig_nw1}$(d)$, but we should note that it already reaches almost $n_w/n_{w,{\rm original}}=10$.
Such a large model will be suffered from a heavy computational burden in terms of both time and storage.
Summarizing above, care should be taken for the choice of parameters contained in AE, which directly affect the number of weights, depending on user's environment, although an increase of number of weights basically leads to acquire a good AE ability.
Especially when an AE includes MLP, we have to be more careful since the fully-connected manner inside MLP leads to the exponential increase for the number of weights.

\subsubsection{Pruning operation}
\label{sec:pru}

%%%%%%%%%%%%%%%%%%%%%%%%%%%%%%%%%%%%%%%%%%%%
\begin{figure}
    \centering
    \includegraphics[width=0.60\textwidth]{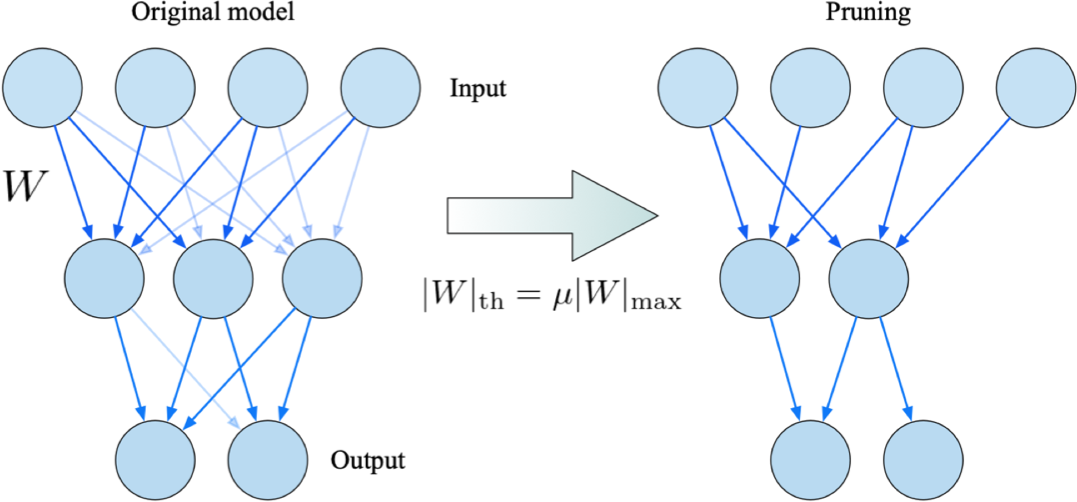}
    \caption{Pruning for multi-layer perceptrons.}
    \label{fig_prusch}
\end{figure}
%%%%%%%%%%%%%%%%%%%%%%%%%%%%%%%%%%%%%%%%%%%%

\begin{table}
\centering
  \caption{Models for the pruning study.}
\begin{tabular}{ccc}
\hline\hline
Examples & \# of latent space $n_r$ & Activation function \\\hline
Cylinder wake  & 16   & Swish   \\
Transient  & 16   & tanh    \\
Sea surface temperature  & 2   & tanh    \\
Turbulent channel flow  & 1024   & Swish \\
\hline\hline
\end{tabular}
\label{tab3}
\end{table}

Through the investigations above, we found that the appropriate parameter choice enables the AE to improve its mapping ability.
Although we may encounter the problem of weight updates as reported above, increasing the number of weights $n_w$ is also essential to achieve a better AE compression.
However, if the AE model could be constructed well with a large number of weights, the increase of $n_w$ also directly corresponds to a computational burden in terms of both time and storage for {\it a posteriori} manner. 
Hence, our next interest here is whether an AE model trained with a large number of weights can maintain its mapping ability with a pruning operation or not.
The pruning is a method to cut edges of connection in neural networks and has been known as a good candidate to reduce the computational burden while keeping the accuracy in both classification and regression tasks~\cite{han2015deep}.
We here assess the possibility to use the pruning for AEs with fluid field data.
As illustrated in figure~\ref{fig_prusch}, we do pruning by taking a threshold based on a maximum value of weights per each layer such that $|W|_{\rm th}=\mu |W|_{\rm max}$, where $W$ and $\mu$ express weights and a sparsity threshold coefficient, respectively.
An original model without pruning corresponds to $\mu=0$.
The pruning for the CNN is also performed as well as MLP.
For this assessment, we consider the best models in our investigation for activation functions (section~\ref{sec:caf}), as summarized in table~\ref{tab3}.

%%%%%%%%%%%%%%%%%%%%%%%%%%%%%%%%%%%%%%%%%%%%
\begin{figure}
    \centering
    \includegraphics[width=0.83\textwidth]{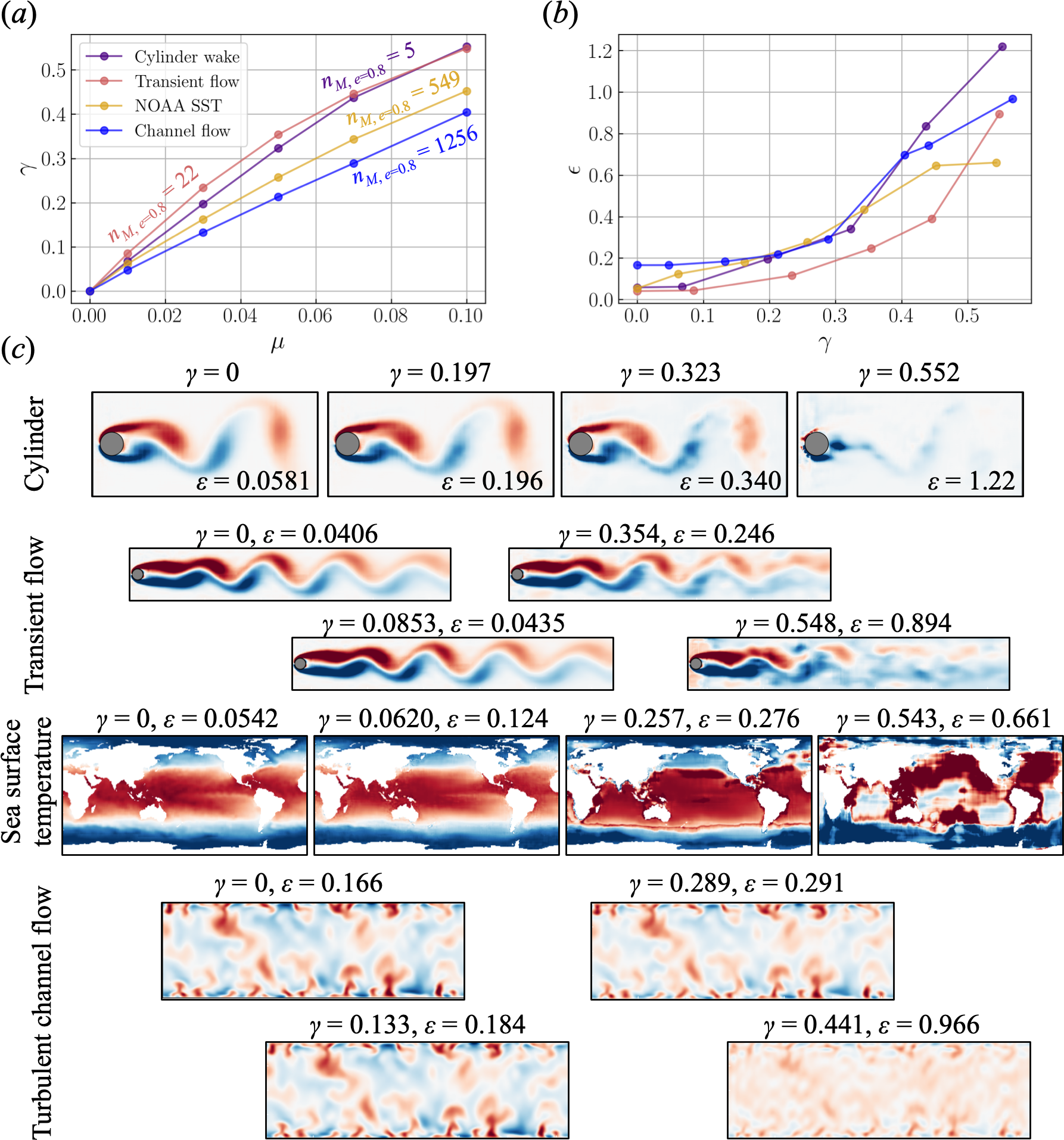}
    \caption{Pruning for AEs with fluid flows. $(a)$ Relationship between 
    %a 
    the
    sparsity threshold coefficient $\mu$ and 
    %a
    the 
    sparsity factor $\gamma=1-n_w/n_{w,{\rm original}}$. The number of singular modes which achieves $80\%$ cumulative energy $n_{M,~e=0.8}$ is also shown. $(b)$ Relationship between a sparsity factor $\gamma$ and $L_2$ error norm $\epsilon$. $(c)$ Representative flow fields of each example.}
    \label{fig_prures}
\end{figure}
%%%%%%%%%%%%%%%%%%%%%%%%%%%%%%%%%%%%%%%%%%%%

Let us present in figure~\ref{fig_prures} the results of pruning operation for the AEs with fluid flows.
To check the weight distribution of AEs for each problem setting, the relationship between the sparsity threshold coefficient $\mu$ and the sparsity factor $\gamma=1-n_w/n_{w,{\rm original}}$ is studied in figure~\ref{fig_prures}$(a)$.
With all cases, $\mu$ and $\gamma$ have a proportional relationship as can be expected.
Noteworthy here is that the sparsity of the model for the turbulent channel flow is lower than that for the others over the considered $\mu$.
This indicates the significant difference in the weight distribution in the AEs in the sense that the contribution of the weights in the turbulence example for reconstruction is distributed more uniformly across the entire system.
In contrast, for the laminar cases, i.e., the periodic shedding and transient flows, a fewer number of weights has the higher magnitudes and a contribution for reconstruction than the turbulence case.
The curve of sea surface temperature model shows the intermediate behavior between them.
It is also striking that the aforementioned trend is analogous to the singular value spectra in figure~\ref{fig_flows}.
The relationship between the sparsity factor $\gamma$ and the $L_2$ error norm with some representative flow fields are shown in figures~\ref{fig_prures}$(b)$ and $(c)$.
As presented, the reasonable recovery can be performed up to approximately 10 to 20\% pruning.
For reader's information, the number of singular modes which achieves $80$\% cumulative energy $n_{M,~e=0.8}$ is presented in figure~\ref{fig_prures}$(a)$.
Users can decide the coefficient $\gamma$ to attain a target error value depending on target flows.
Summarizing above, the efficient order reduction by AE can be carried out by considering the considerable methods and parameters as introduced through the present paper.

\section{Concluding remarks}
\label{sec:conclusion}

We presented the assessment of neural network based model order reduction method, i.e., autoencoder (AE), for fluid flows.
The present AE which comprises of a convolutional neural network and multi-layer perceptrons was applied to four data sets, i.e., two-dimensional cylinder wake, transient process, NOAA sea surface temperature, and turbulent channel flow.
The model was evaluated in terms of various considerable parameters in deciding a construction of AE.
With regard to the first investigation for the number of latent modes, it was clearly seen that the mapping ability of AE highly depends on the target flows, i.e., complexity.
In addition, the first assessment enables us to notice the importance of investigation for the choice of activation function and the effect of number of weights contained in the AE.
Motivated by the first assessment, the choice of activation function was then considered.
We found that we need to be careful for the choice of 
of activation functions at hidden layers of AE so as to achieve the effective order reduction because the influence of activation functions highly depends on target flows.
At last, the dependence of the reconstruction accuracy by the AE on the number of weights contained in the model was assessed.
We exhibited that care should be taken to change the amount of parameters of AE models because we may encounter the problem of weight updates by using many numbers of weights.
The use of pruning was also examined as one of the considerable candidates to reduce the number of weights.
Although it has been known that the capability of neural networks may be improved by re-training of pruned neural networks~\cite{han2015learning}, the possibility of this strategy will be tackled in future.

{We should also discuss remaining issues and considerable outlooks regarding the present form of AE.
One of them is the interpretability of the latent space, which may be regarded as one of the weaknesses of neural network-based model order reduction due to its black-box characteristics.
Since the AE-based modes are not orthogonal with each other, it is still hard to understand the role of each latent vector for reconstruction~\cite{omata2019}.
To tackle this issue, Murata et al.~\cite{MFF2019} proposed a customized CNN-AE which can visualize individual modes obtained by the AE, considering a laminar cylinder wake and its transient process.
However, they also reported that the applications to flows requiring a lot of spatial modes for their representation, e.g., turbulence, are still challenging with their formulation.
This is because the structure of the proposed AE would be more complicated with increasing the complexity of target problems, and this eventually leads to difficulty in terms of interpretability.
The difficulty in applications to turbulence could also be found in this paper, as discussed above.
More recently, Fukami et al.~\cite{FNF2020} have attempted to use a concept of hierarchical AE to handle turbulent flows efficiently, although it still needs further improvement for handling turbulent flows in a sufficiently interpretable manner.
Although the aforementioned challenges are just examples, through such continuous efforts, we hope to establish in near future more efficient and elegant model order reduction with additional AE designs.

Furthermore, of particular interest here is how we can control a high-dimensional and nonlinear dynamical system by using the capability of AE demonstrated in the present study, which can promote efficient order reduction of data thanks to nonlinear operations inside neural networks. 
From this viewpoint, it is widely known that capturing appropriate coordinates capitalizing on such model order reduction techniques including AE considered in the present study is also important since being able to describe dynamics on a proper coordinate system enables us to apply a broad variety of modern control theory in a computationally efficient manner while keeping its interpretability and generaliziability~\cite{fukami2020sparse}.
To that end, there are several studies to enforce linearity in the latent space of AE~\cite{lusch2018deep,champion2019data}, although their demonstrations are still limited to low-dimensional problems that are distanced from practical applications.
We expect that the knowledge obtained through the present study, which addresses complex fluid flow problems, can be transferred to such linear-enforced AE techniques towards more efficient control of nonlinear dynamical systems in near future.
}

\section*{Acknowledgement}
This work was supported from Japan Society for the Promotion of Science (KAKENHI grant number: 18H03758, {21H05007}).

\section*{Declaration of interest}

The authors report no conflict of interest.

\bibliographystyle{sp_fixed}      % mathematics and physical sciences
\bibliography{references}   % name your BibTeX data base

\end{document}